\documentclass[iop]{emulateapj}

\slugcomment{}

\shorttitle{NGC 3516}
\shortauthors{Devereux et al.}

\begin{document}
\columnsep=8.5mm
\textwidth=18.55cm
\title{Photoionization Modeling of the Low Luminosity Seyfert 1 Nucleus in NGC 3516}

\author{Nick Devereux}
\affil{Department of Physics \& Astronomy, Embry-Riddle Aeronautical University,
         Prescott, AZ 86301: devereux@erau.edu}

\begin{abstract}

Spectroscopic observations of the low luminosity Seyfert 1 nucleus in NGC 3516 obtained with the Hubble Space Telescope show that the visible spectrum is dominated by the Balmer emission lines of Hydrogen (H)
and a continuum luminosity that rises into the UV.  The 
anomalous H${\alpha}$/H${\beta}$ emission line ratio, the Balmer emission line luminosity and the 
distinctive shape observed for the H${\alpha}$ emission line profile serve as important constraints in any photoionization model aimed at explaining the visible emission line spectrum of NGC 3516.
Photoionization modeling using Cloudy demonstrates that the central UV--X-ray source is able to completely ionize the H gas in between the Balmer and dust reverberation radii if the electron density is ${\le}$ 3 ${\times}$ 10${^7}$ cm${^{-3}}$ throughout. Thus, according to this model
the region responsible for producing the visible H lines is a dust free shell of ionized H gas.
Interestingly, the model predicts a rapid rise in the electron temperature as the central UV--X-ray source is approached, mirrored by an equally precipitous decrease in the Balmer line emissivity that coincides
with the Balmer reverberation radius, providing a natural explanation for the finite width observed for the H Balmer lines. Collectively, the merit of the model is that it explains the relative intensities of the 
three brightest Balmer lines, and the shape of the H${\alpha}$ emission line profile. However, questions remain concerning the unusually weak forbidden lines that can not be addressed using Cloudy due to limitations with the code.

\end{abstract}


\section{Introduction}

The visible spectrum of the nucleus of NGC 3516 was first described by \cite{Sey43} and includes bright, and unusually broad permitted Balmer emission lines of hydrogen (H), narrow [O\,{\sc iii}]${\lambda\lambda}$4959,5007, and absent  [O\,{\sc ii}]${\lambda}$3727 forbidden emission lines. Collectively, these spectroscopic features have become a defining characteristic of Seyfert 1 active galactic
nuclei (AGN) and much effort has been expended since in trying to understand their origin as reviewed recently by \cite{Ho08}.

NGC 3516 is of potentially great importance in deciphering the mystery surrounding the origin of broad H emission lines because this low luminosity active galactic nucleus (LLAGN)
is also time-variable allowing various independent measures of the broad line region (BLR) size. For example, correlated variability between the H${\beta}$ emission line and the adjacent continuum lead to a Balmer reverberation lag ${\sim}$ 7  l.d. \citep{Den10}. However, complementary observations of correlated time variability between the visible and near infrared continua point to a 2.2\,${\micron}$ ($K$-band) dust reverberation radius of ${\sim}$ 50 - 70 l.d  \citep{Kos14}. Thus, the dust sublimation radius is 
about an order of magnitude larger
than the Balmer reverberation radius. 
This size discrepancy is very significant and is evidently a common feature of reverberating AGN \citep{Kos14} begging the question 
{\it what lies in between?}
A plausible answer to this question is the model proposed by \cite{Net93},
in which the central UV--X-ray source is able to sublimate dust from a sizeable volume of H gas, permitting 
it to be photoionized. According to this interpretation the Balmer reverberation radius marks just the inner radius of a much larger volume of photoionized gas, an insight articulated previously by \citet[][and references therein]{Kos14}. In this context it is of interest to note that Seyfert alikened the visible spectrum of NGC 3516 to that of a planetary nebula which
bears some geometrical resemblence to the BLR photoionization model of \cite{Net93}. 

\begin{deluxetable*}{lcccccccl}
\tabletypesize{\scriptsize}
\tablecolumns{9} 
\tablecaption{NGC 3516 STIS Datasets\label{tbl-2}}
\tablewidth{0pt}
\tablehead{
\colhead{PID} & \colhead{Observation Date} & \colhead{Grating} & \colhead{Spectral Range} & \colhead{Slit} & \colhead{Dispersion} & \colhead{Plate Scale} & \colhead{Integration Time} & \colhead{Datasets}   \\
\colhead{} & \colhead{} &  \colhead{} & \colhead{\AA}  & \colhead{arc sec} & \colhead{\AA/pixel}& \colhead{arc sec/pixel} & \colhead{s} & \colhead{}\\
\colhead{(1)} & \colhead{(2)} &  \colhead{(3)} & \colhead{(4)}  & \colhead{(5)} & \colhead{(6)} & \colhead{(7)} & \colhead{(8)} & \colhead{(9)} \\
}
\startdata
7355 & 4-13-1998 &  G140L  & 1150 -- 1730 & 52 x 0.5 & 0.6 & 0.0246 & 32820 & o4st01010 \\
& & & & & & & & --o4st01070\\
7355 & 4-13-1998 &  G430L  & 2900 -- 5700 & 52 x 0.5 & 2.73 & 0.05 & 122891 & o4st02010 \\
 & & & & & & & & -- o4st13030\tablenotemark{a}  \\
8055 & 6-18-2000 &  G750M  & 6295 -- 6867 & 52 x 0.2 & 0.56 & 0.05 & 1956 & o56c01020 -- \\
& & & & & & & & -- o56c01030 \\
8055 & 6-18-2000 &  G750M  & 6295 -- 6867 & 52 x 0.1 & 0.56 & 0.05 &  60 & o56c01040  \\
8055 & 6-18-2000 & G430L & 2900 -- 5700 & 52 x 0.2 & 2.73 & 0.05 & 600 & o56c01050 \\
\enddata

\tablenotetext{a}{Omitting o4st06030, o4st06040, o4st07030, o4st11020, o4st11030}

\end{deluxetable*}

\begin{figure*}
\epsscale{1.0}
\begin{center}
 \includegraphics[width=130mm,clip]{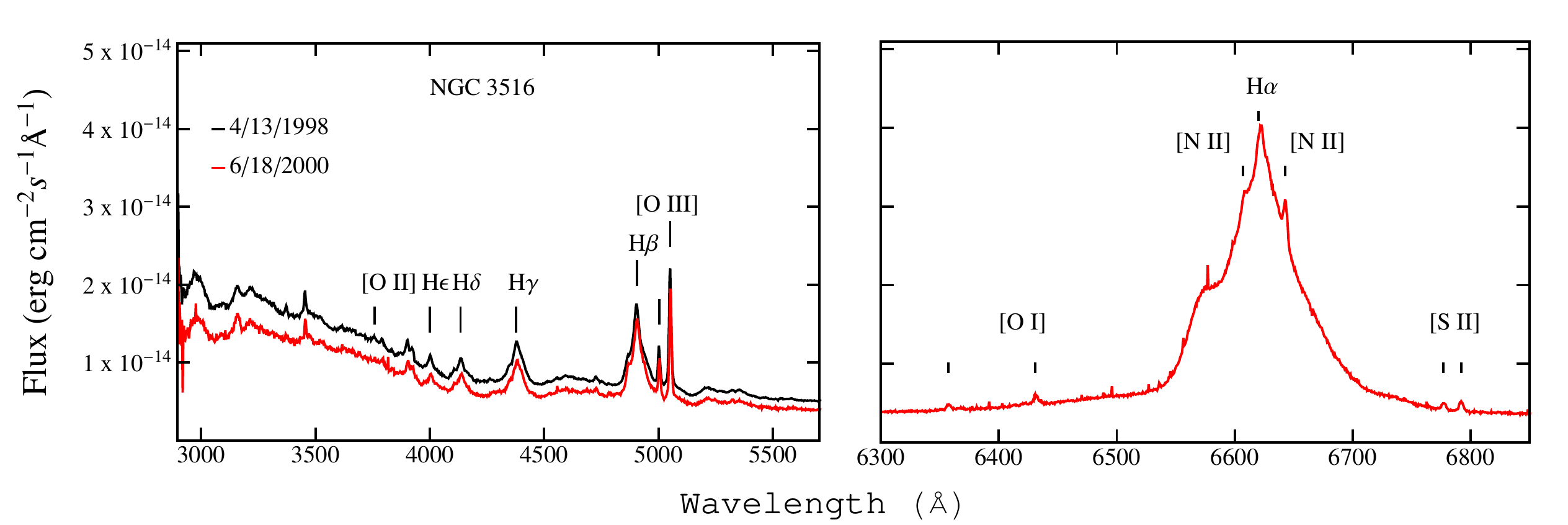}
 \caption{Visible spectra of NGC 3516 as seen through the following gratings: Left panel: G430L. Right panel: G750M. The black line shows data obtained under PID 7355. The red (lighter shade) line for both panels shows data obtained under PID 8055.}
\label{default}  
\end{center}
\end{figure*}

Of all telescopes, the {\it Hubble Space Telescope} ({\it HST}) provides the clearest view 
of the bright AGN in NGC 3516. Consequently, the main objective of this paper 
is to interpret the exquisite visible spectra obtained with {\it HST}  in the context of the \cite{Net93} photoionization model. An important constraint 
in any such model is the shape and amplitude of the ionizing continuum.
Recent observations with ${\it XMM-Newton}$ allowed \cite{Vas09} to constrain the ionizing continuum of the AGN in NGC 3516 as the combination of emission from an accretion disk and an X-ray power law. That ionizing continuum is adopted here as an input to the photoionization code Cloudy \citep{Fer13}
which can predict the relative intensity of the Balmer emission lines for 
various radial density distributions of photoionized gas.
Additionally, since the central black hole (BH) mass is known
\citep{Den10} the shape of the 
broad Balmer emission lines can be used to constrain the Balmer emission line emissivity given a kinematic description for the gas. When combined with the X-ray luminosity, the BH mass implies that the AGN in NGC 3516 is radiating at ${\sim}$ 0.6\% the Eddington luminosity limit \citep{Vas09} and is therefore unable to sustain a radiatively driven outflow. Thus, the BLR gas kinematics are most likely dominated 
by gravity. Time resolved spectra discussed by \cite{Den09} indicate that the BLR is actually an inflow of H gas.
Paradoxically, the AGN in NGC 3516 may also be associated with an outflow \citep[][and references therein]{Bar09} driven by two jets,
the orientation and geometry of which has been discussed previously by \citet{Fer98}, such jets are comprised of a relativistic plasma producing extended radio continuum and collisionally excited forbidden line emission, but little or no Balmer emission. Consequently, the broad Balmer emission lines most likely 
originate in H gas that is photoionized by the central UV--X-ray source. The main objective of this paper is to test that conjecture.

The layout of the paper is as follows. A review of the UV and visible spectra of NGC 3516 obtained with
the Space Telescope Imaging Spectrograph (STIS)
is presented in Section 2. These observations are combined with a model for the ionizing continuum  presented 
in Section 2.2.  Emission line ratios, corrected for dust extinction, constrain a Cloudy photoionization model for the BLR in NGC 3516 as described in Section 3. 
A discussion follows in Section 4 and Conclusions are presented in Section 5.

\section{Results}

The {\it HST}/STIS observations, described in more detail in the following, provide fluxes and relative intensities
for the H Balmer emission lines along with a measure of their line profile shapes. These observational results provide
key constraints in a photoionization model for the BLR in NGC 3516 that is presented in Section 3.

\subsection{${\it HST/STIS}$ Observations}

NGC 3516 has been visited twice with STIS. First in 1998 when it was observed intensively for a period of two days using the G430L and G140L gratings, then a second short visit just over two years later, in 2000,
when it was observed again with the G430L grating and also the G750M grating. The STIS observations for both visits are summarized in Table 1, and the visible spectra are presented in Figure 1. Some details of those observations have been reported previously by \citet{Edl00} and \citet{Bal14}. A thorough analysis of the UV 
emission line spectrum of NGC 3516 has been presented previously by \cite{Goa99a,Goa99b}.

\begin{figure}
\epsscale{1.0}
 \includegraphics[width=74mm,clip]{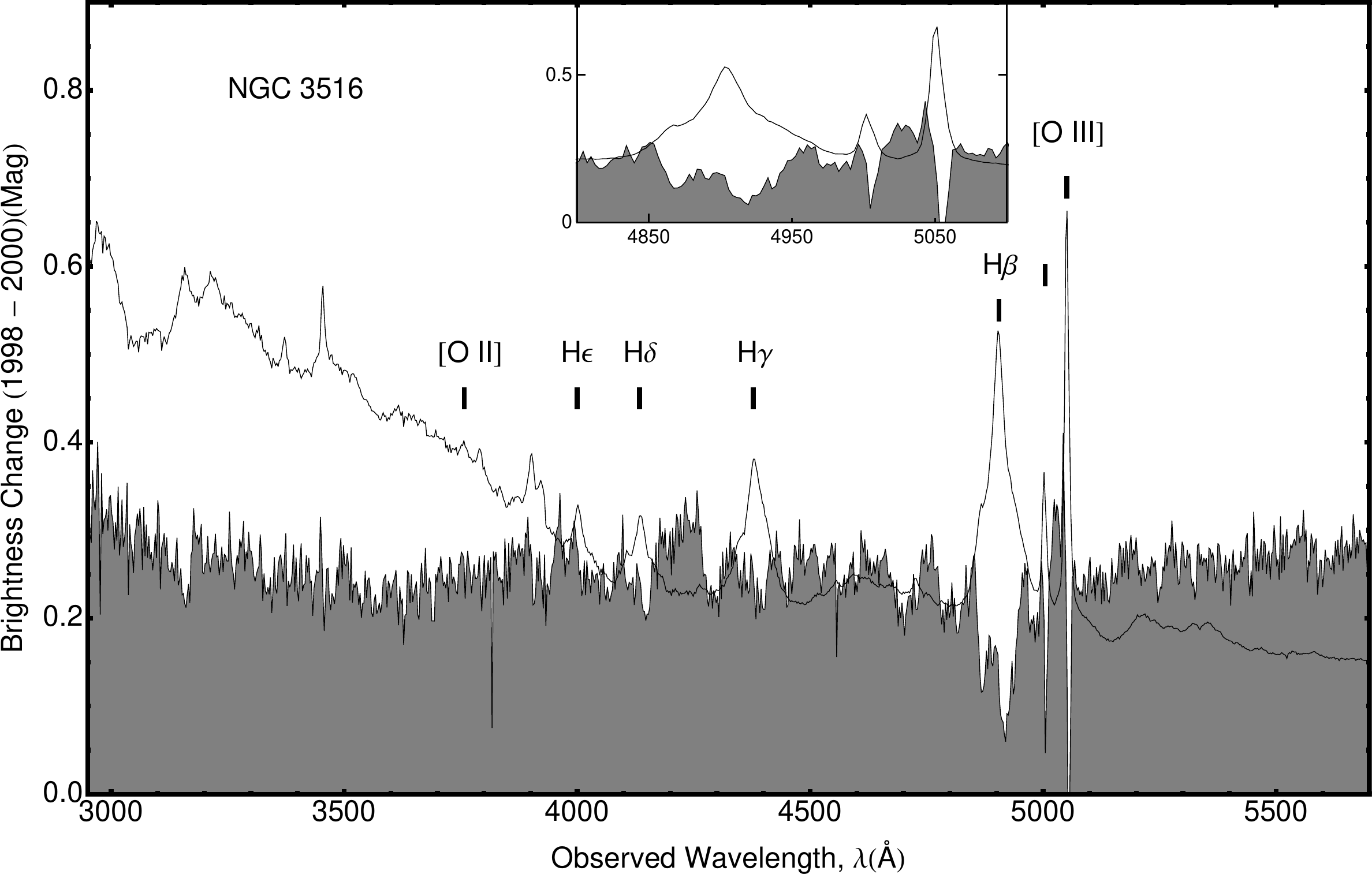}
 \caption{Brightness change between the years 1998 and 2000 as observed with the G430L grating (shaded area). The ordinate indicates the brightness change in magnitudes, the abscissa is wavelength in Angstroms. The 1998 G430L spectrum (black line) is overplotted to illustrate that the flux in the core of the H${\beta}$ emission line changed less than the flux in the wings. The inset shows an expanded region around the H${\beta}$ emission line.}
\end{figure}

Multiple calibrated exposures obtained through each of the G750M and G430L gratings were shifted and combined 
for each grating but separately for each visit using the STSDAS task occreject. Subsequently, 
emission line fluxes were measured using the STSDAS contributed task specfit.
Between the years 1998, and 2000 the visible continuum measured with the G430L grating decreased quite conspicuously by ${\sim}$ 20\% as illustrated in Figure 2.  The decrease in continuum brightness may be caused in part by the smaller slit size employed for the year 2000 observation. However, NGC 3516 is 
also known to be reverberating \citep{Den10} and the ${\sim}$ 5\% decrease in the H${\beta}$ emission line flux between the two observations is consistent with prior observations. Intriguingly, the flux in the core of the H${\beta}$ emission line changed less than the flux in the wings, as the inset to Figure 2 shows in more detail.

The Balmer series of H, and the [O\,{\sc iii}]${\lambda\lambda}$4959,5007 forbidden emission lines dominate the visible emission line spectrum. A small spike on the red side of the broad H${\alpha}$ emission line profile coincides with the vacuum wavelength 6585.28 {\AA} expected for the brightest [N\,{\sc ii}] forbidden emission line.  A model for the brighter [N\,{\sc ii}] line was constructed that, when removed, did not oversubtract the broad H${\alpha}$ emission line profile which is otherwise smoothly varying. Although it can not be seen, the fainter vacuum wavelength 6549.85 {\AA}  [N\,{\sc ii}] emission line is constrained by atomic physics to have the same width and one third the flux of the brighter line.  Line fluxes are reported in Table 2 for all the emission lines seen in the G750M spectrum including the broad H${\alpha}$ emission line, the [N\,{\sc ii}] forbidden emission lines, the density sensitive [S\,{\sc ii}] vacuum wavelength 6718.29 {\AA}, and 6732.67 {\AA} lines, plus the two [O\,{\sc i}] vacuum wavelength 6302.04 {\AA}, and 6365.53 {\AA} lines. 

The Balmer series of H continues into the G430L spectrum. Emission line fluxes are reported in Table 2 for  H${\beta}$, H${\gamma}$, H${\delta}$ and H${\epsilon}$. The latter two lines are considerably fainter, and the adjacent continuum is not flat which introduces an additional model dependent systematic uncertainty, due to the continuum subtraction, that is difficult to quantify.
Fluxes are also reported for the vacuum wavelength 4960.30 {\AA} and 5008.24 {\AA}  [O\,{\sc iii}] emission lines.  An 
upper limit is reported in Table 2 for the vacuum wavelength 4364.44 {\AA}  [O\,{\sc iii}] emission line which is overwhelmed by the broad H${\gamma}$ line. A flux is also reported for the unresolved vacuum wavelength 3727.09, 3729.88 {\AA}  [O\,{\sc ii}] doublet.

\begin{deluxetable}{lccc}
\tabletypesize{\footnotesize}
\tablecaption{Emission Line Parameters for the Combined G430L Nuclear Spectra Obtained 4-13-1998
and the Combined G750M Nuclear Spectrum Obtained 6--18--2000\tablenotemark{a}}
\tablewidth{0pt}
\tablehead{
\colhead{Line} & \colhead{Central\tablenotemark{b}} & \colhead{Flux\tablenotemark{c}} & \colhead{FWHM}   \\
\colhead{} & \colhead{Wavelength (\AA)} &  \colhead{(10$^{-14}$ erg cm$^{-2}$ s$^{-1}$)} & \colhead{(kms$^{-1}$)} \\
\colhead{(1)} & \colhead{(2)} &  \colhead{(3)} & \colhead{(4)} \\
}
\startdata
$\textrm{[O\,\sc ii]}$  & 3755 ${\pm}$ 1 &  0.8 ${\pm}$ 0.1 & 1255 ${\pm}$ 240 \\
H${\epsilon}$ (broad) & 4000 &  2.9 ${\pm}$ 0.1 & ... \\
H${\delta}$ (broad) & 4134 &  12.5 ${\pm}$ 0.1 & ... \\
H${\gamma}$ (broad) & 4378 &  30 ${\pm}$ 0.1 & 3323 ${\pm}$ 180 \\
$\textrm{[O\,\sc iii}]$\tablenotemark{d} &  4401  &  ${\leq}$ 0.3  &  1000 \\
H${\beta}$ (broad) & 4905  &  61 ${\pm}$ 0.1 & 2540 ${\pm}$ 170 \\
H${\beta}$ (broad)\tablenotemark{e} & 4905  &  58 ${\pm}$ 0.1 & 2540 ${\pm}$ 170 \\
$\textrm{[O\,\sc iii}]$ &  5004  ${\pm}$ 1 & 7 ${\pm}$ 0.1 &  910 ${\pm}$ 224 \\
$\textrm{[O\,\sc iii}]$ &  5051  ${\pm}$ 1 & 20 ${\pm}$ 0.1 &  1200 ${\pm}$  40 \\
$\textrm{[O\,\sc i]}$ &  6357  ${\pm}$ 1 & 0.3 ${\pm}$ 0.1 &  203 ${\pm}$  29 \\
$\textrm{[O\,\sc i]}$  &  6431  ${\pm}$ 1 & 0.4 ${\pm}$ 0.1 &  200\\
$\textrm{[N\,\sc ii]}$ &  6607 ${\pm}$ 3   & 0.83  &  200   \\
H${\alpha}$ (broad) & 6620  & 307 ${\pm}$ 0.1 & 2682 ${\pm}$ 60 \\
$\textrm{[N\,\sc ii]}$ & 6643 ${\pm}$ 0.3  & 2.5\tablenotemark{f} & 200   \\
$\textrm{[S\,\sc ii]}$ & 6777 ${\pm}$ 3  & 0.5 ${\pm}$ 0.2  & 250 ${\pm}$ 100 \\
$\textrm{[S\,\sc ii]}$ & 6792 ${\pm}$ 7 & 0.7    & 250 \\

\enddata
\tablenotetext{a}{Table entries that do not include uncertainties are fixed parameters. }
\tablenotetext{b}{Observed wavelength}
\tablenotetext{c}{4-13-1998 observations measured within
a 0.5{\arcsec}  x 0.35{\arcsec}  aperture. 6-18-2000 observation, measured within
a 0.2{\arcsec}  x 0.35{\arcsec}  aperture. Continuum subtracted but not corrected for dust extinction. Model dependent systematic uncertainties introduce an additional ${\sim}$3\% error not reported in the Table.}
\tablenotetext{d} {The [O III] emission line parameters chosen so as to not over-subtract the broad H${\gamma}$ emission line profile}
\tablenotetext{e} {6-18-2000 observation}
\tablenotetext{f} {The [N II] emission line flux is chosen so as to not over-subtract the broad H${\alpha}$ emission line profile. }

\end{deluxetable}

Collectively, the {\it HST} spectra bear an uncanny resemblence to the one described by \cite{Sey43}.
Using photographic plates he measured the relative intensities of the Balmer lines and the [O\,{\sc iii}]${\lambda\lambda}$4959,5007 forbidden emission lines, to be within 25\% of the values measured with STIS. However, an inconsistency has been found with \cite{Edl00} who report H${\beta}$ and H${\gamma}$ emission line fluxes that are one order of magnitude larger than cited in Table 2. Including the forbidden [N\,{\sc ii}] emission lines, the broad H${\alpha}$ flux reported by \cite{Bal14} agrees with the value cited in Table 2 within the 3\% uncertainty expected for plausible, but different models of the underlying continuum. 

\begin{figure}
\epsscale{1.0}
 \includegraphics[width=84mm,clip]{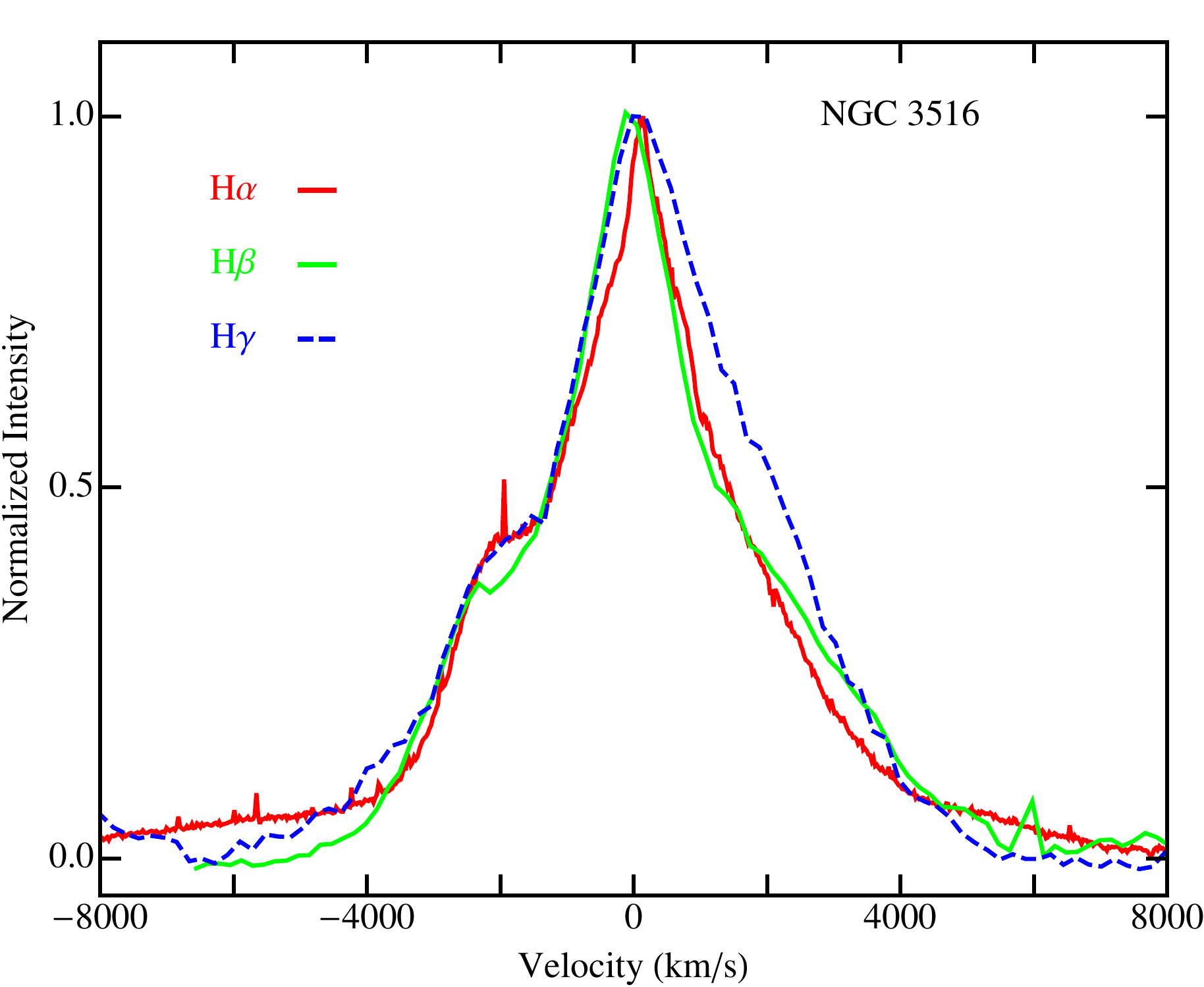}
 \caption{Normalized H${\alpha}$ (red), H${\beta}$ (green), and H${\gamma}$ (blue-dashed) emission lines profiles  plotted as a function of rest-frame velocity.  }
\end{figure}

The broad H${\alpha}$ emission line has a single peak, but is obviously asymmetric due to a ``bump" on the blue side of the profile illustrated in Figure 3. This feature was seen, and commented on previously by \cite{Bok77}, \cite{Wan93}
and \cite{Pop02}. Evidently, the feature is real, and has persisted for at least 25 years. Adopting a heliocentric recession velocity of 2508 ${\pm}$ 60 km/s for NGC 3516, measured using the peak of the brightest [O\,{\sc iii}] emission line (Figure 1), allows wavelength to be converted into rest frame velocity for each of the H${\alpha}$, H${\beta}$, and H${\gamma}$ emission line profiles.  Figure 3 illustrates that the emission line profile shapes are very similar to each other after they have been normalized to their respective peak intensity. The ``bump" on the blue side is seen in H${\alpha}$, H${\beta}$, and H${\gamma}$. The fact that the Balmer emission line profiles are so similar suggests that the dust extinction in the visible part of the spectrum, internal to the BLR, is essentially zero.

Using the results provided in Table 2, observed ratios involving the fluxes for the three brightest Balmer lines; H${\alpha}$/H${\beta}$, and H${\beta}$/H${\gamma}$, are reported in Table 3 along with the canonical Case B values expected for an
idealized nebula of uniform electron temperature, corresponding to 10${^4}$ K, and a uniform electron density of 10${^4}$ cm${^{-3}}$ \citep{Hum87}.
Interestingly, the H${\alpha}$/H${\beta}$ ratio measured with STIS agrees with the average of the values reported previously in \cite{Bok77}, and
is almost a factor of two larger than the Case B value. Such deviations from recombination theory have been noted for other AGN \cite[][and references therein]{Dev13} and can be explained in terms of collisional excitation, enhancing just H${\alpha}$, relative to the other Balmer lines. However, the Balmer emission line ratios can also be affected by dust extinction which is 
addressed in the next section.

\begin{deluxetable}{ccccc}
\tabletypesize{\footnotesize}
\tablecaption{Observed Emission lines Compared to Case B, and Cloudy Model Predictions}
\tablewidth{0pt}
\tablehead{
\colhead{Ratio} & \colhead{Observed} & \colhead{Extinction} & \colhead{Case B{\tablenotemark{a}}}  & \colhead{Cloudy}  \\
 & & Corrected &  &  \\
\colhead{(1)} & \colhead{(2)} &  \colhead{(3)}  & \colhead{(4)} & \colhead{(5)} \\
}
\startdata
H${\alpha}$/H${\beta}$ & 5.2 ${\pm}$ 0.01 & 5.0 ${\pm}$ 0.01 &  2.8    & 5.2 \\
H${\beta}$/H${\gamma}$ & 2.0 ${\pm}$ 0.01 & 2.0 ${\pm}$ 0.01 & 2.1   & 1.9  \\
$\textrm{[O\,\sc iii}]$${\lambda}$5007/H${\beta}$ & 0.31 ${\pm}$ 0.02 & 0.30 ${\pm}$ 0.02& ... &  0.37{\tablenotemark{b}} \\
$\textrm{[O\,\sc iii}]$${\lambda}$4959/H${\beta}$ & 0.10 ${\pm}$ 0.02 & 0.10 ${\pm}$ 0.02 &... &  0.12{\tablenotemark{b}} \\
\enddata
\tablenotetext{a}{Assuming a uniform electron temperature of 10${^4}$ K, and a uniform electron density of 10${^4}$ cm${^{-3}}$.}
\tablenotetext{b}{${\frac{1}{10}}$ solar metallicity. See Table 4.}
\end{deluxetable}

\begin{figure}
\epsscale{1.0}
\includegraphics[width=84mm,clip]{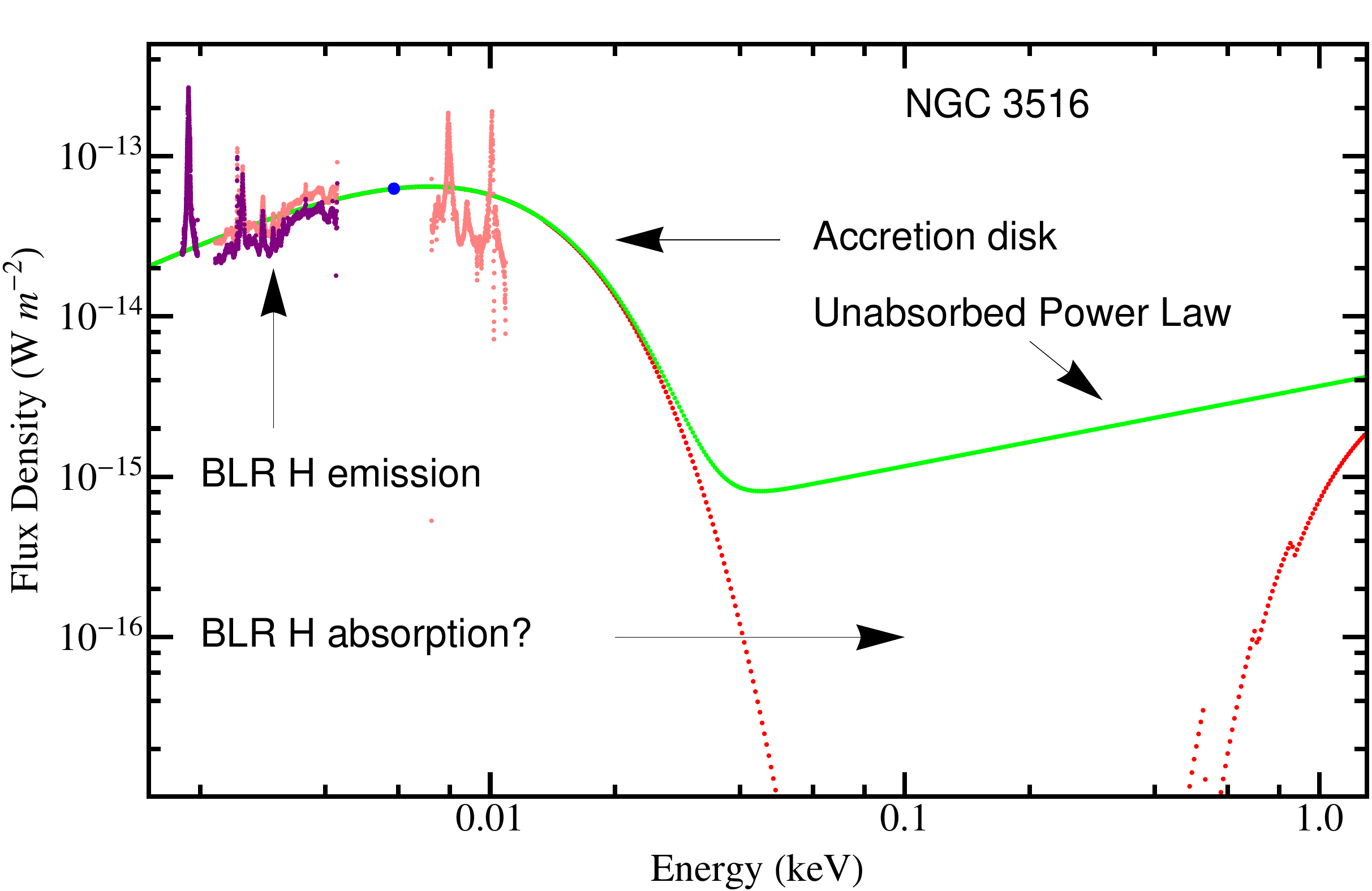}
\caption{The visible--UV--X-ray continuum of NGC 3516. The solid green line is the model unabsorbed continuum, the dotted red line is a model representation of H photoelectric absorption. Data for both lines, plus the single
XMM-OM measurement represented by the blue dot, courtesy of \cite{Vas09}.
STIS spectra obtained under PID 7355 are plotted in pink (lighter shade), and PID 8055 plotted in purple (darker shade).}
\end{figure}

\subsection{UV--X-ray Continuum, and Foreground Dust Extinction}

\cite{Vas09} modeled the UV--X-ray continuum of NGC 3516 in terms of a blackbody, representing an accretion disk, and a power law. The amplitude of that continuum model, recapitulated in Figure 4, is constrained by a single XMM-OM observation obtained at the end of the year 2001. However, 
the model continuum agrees with that measured in the G430L spectrum  and the mean extinction corrected 1365${\rm \AA}$ continuum discussed previously by \cite{Goa99b}. Consequently, the model continuum presented in Figure 4 provides a useful constraint on the production rate of H ionizing photons by the central UV--X-ray source. 
For a distance of 38 Mpc (R.B. Tully, private communication) numerically integrating the continuum yields 1.2 ${\times}$ 10$^{53}$ H ionizing photons s$^{-1}$, of which the majority, ${\sim}$ 75\%, are produced by the accretion disk\footnotemark[1] \footnotetext[1]{ Characterized by the parameter ${\it T_{max}}$ = 0.00288 keV. See \cite{Vas09} for details.} and the remainder by the power law. These results constrain a photoionization model for the BLR in NGC 3516 discussed further in Section 3.

As illustrated in Figure 4, the G430L continuum measured in 1998 coincides almost identically with the model accretion disk, but the contemporaneous G140L spectrum lies significantly below. Although one can not rule out time variability
as the reason for the discrepancy, the likelihood that the observed continuum is representative of the mean 
provides an opportunity to estimate the dust extinction by comparison with the model continuum. 
If one assumes a Galactic form for the redenning law, A$\rm_{v}$/E(B-V) = 3.2 \citep{Car89} then a least squares analysis on the difference between the model continuum and the continuum measured in the contemporaneous G430L and G140L spectra yields the following significant result, 

\begin{equation}
A_{\lambda} =\frac{(1163 \pm 95)}{\lambda(\rm{\AA})} -0.06~~~~\rm{mag}
\end{equation}

which predicts a color excess E(B-V) = 0.05 ${\pm}$ 0.01 consistent with the range of values for the Galactic extinction quoted by \cite{Goa99b}. Consequently, the foreground extinction towards NGC 3516 
at the wavelength of the H${\alpha}$ emission line is likely to be small, ${\sim}$ 0.1 mag. Values for the ratios H${\alpha}$/H${\beta}$, and H${\beta}$/H${\gamma}$, are reported in Table 3 corrected
for Galactic extinction using equation 1.

\section{Photoionization Modeling of the BLR using Cloudy}

Evidently, the reason that the LLAGN in NGC 3516 is so bright is because the foreground visible dust extinction is essentially zero.  Furthermore, the dust extinction internal to the BLR may also be
zero since the Balmer emission line profile shapes are so similar (see Figure 3). Thus, the AGN is essentially completely exposed allowing a very clear view of the BLR. This
is perhaps not entirely unexpected as \cite{Kos14} measure the dust reverberation radius to be significantly larger than the Balmer reverberation radius \citep{Den10}.
Collectively, the {\it HST}/STIS observations suggest a model for the BLR of NGC 3516, advocated previously by \cite{Net93},
in which the central UV--X-ray source is able to sublimate dust from a sizeable volume of H gas, permitting 
it to be photoionized. The implications of such a model are explored in the following using version 13.02 of Cloudy \citep{Fer13}.

\begin{deluxetable}{l}
\tabletypesize{\scriptsize}
\tablecaption{Input Parameters for the Cloudy Photoionization model of NGC 3516}
\tablewidth{0pt}
\tablehead{
\colhead{Parameter}   \\
}
\startdata

AGN T=8.1e4 K, $\alpha_{\rm{ox}}$=-1.4, $\alpha_{\rm{uv}}$=-0.5, $\alpha_{\rm{x}}$=-0.5\\
q(h)=53.09 \\
cosmic rays background    \\
radius 16.     \\
hden 7.4, power =-0.5 \\
sphere \\
abundances ISM no grains \\
Stop radius 17.35  \\
iterations 2 \\
\enddata
\end{deluxetable}

\begin{figure}
\includegraphics[width=84mm,clip]{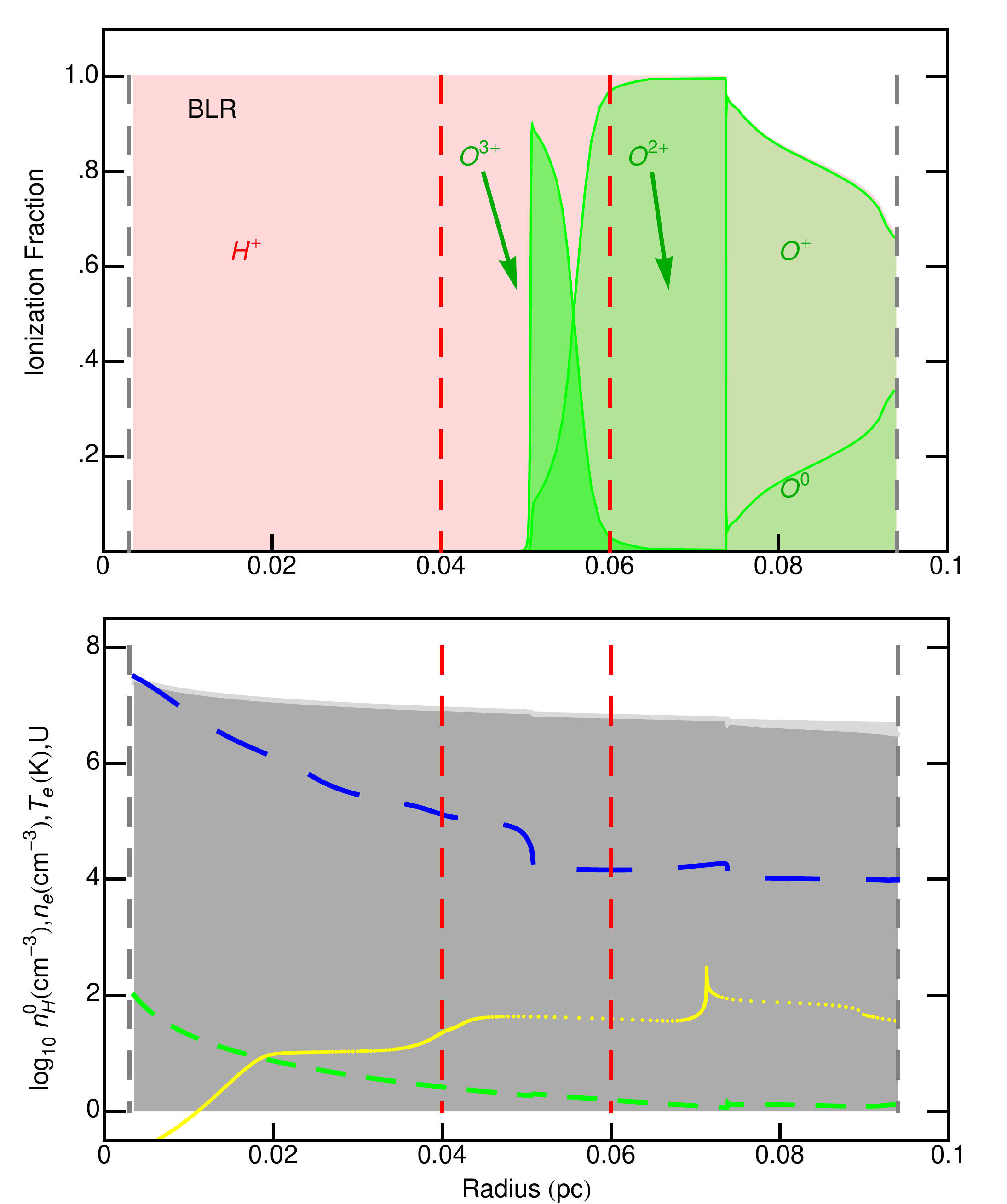}
\caption{Radial distribution of various Cloudy model results. Radial distance from the central BH 
in pc is indicated on the abscissa. The ordinate refers to various units. (Upper panel) Cloudy model radial distributions of the H and O ionization fraction. (Lower panel) Neutral H density (light grey shading), electron density (dark grey shading), elecron temperature (long-dashed blue line), logarithm to the base 10 of the H${\alpha}$ emissivity in arbitrary units (dotted yellow line) and ionization parameter (short-dashed green line). The two inner vertical dashed red lines in both plots represent the range of dust reverberation radii measured by \cite{Kos14}.  The two outer vertical dashed grey lines
identify the inner and outer radius of the model (see Table 4), the latter
corresponding also to the luminosity radius defined in Section 3.2}
\end{figure}

Table 4 summarizes the parameters 
employed to model photoionization of the BLR in NGC 3516 a full description of which can be found in the Cloudy documentation. Briefly, they describe a spherically symmetric distribution of neutral H gas that is photoionized by the central UV--X-ray source.  

The radial number density distribution for the neutral gas is represented by an r${^{-n}}$, power law, normalized by a number density at the inner radius, ${\rho}$. A grid of photoionization models 
spanning
0 ${\leq}$  n ${\leq}$ 1.5 and  7.0 ${\leq}$ log${_{10}}$${\rho}$(cm${^{-3}}$) ${\leq}$ 8.0 was constructed
in order to discover the intersection of model predictions for the intrinsic H${\alpha}$/H${\beta}$, H${\beta}$/H${\gamma}$, and [O\,{\sc iii}]${\lambda\lambda}$4959,5007/H${\beta}$
emission line ratios with the extinction corrected
values. Subsequent optimization of the density versus outer radius yielded emission
line ratios that are within ${\sim}$ 5\% of the observed extinction corrected values reported
in Table 3.  As explained in more detail in the following sections, the modeling results point to low density and possibly low metallicity gas as the origin of the visible emission line spectrum
observed for NGC 3516.

\subsection{Radial Structure, and Physical Properties of the BLR in NGC 3516}

Of particular interest in understanding the physical conditions that may exist in the BLR is what the photoionization code Cloudy has to say about the radial distributions of the ionization fraction,
the electron density, the electron temperature, the H${\alpha}$ emission line emissivity,
and the ionization parameter. These results, depicted in Figure 5, represent the model parameters listed 
in Table 4. Some interesting trends are apparent.
First, the upper panel in Figure 5 shows that the H ionization fraction is
predicted to be 100\% inside the dust reverberation radius measured by \cite{Kos14}. 
Plus, a significant ionization gradient 
is predicted, in the sense that O$^{2+}$ is inevitably ionized to O$^{3+}$, as the central UV-X-ray source is approached. Consequently, the H and O emitting regions are spatially disparate, the 
H emission being produced in a dust-free shell, surrounded by a potentially dusty O emitting region. 
Second, the lower panel of Figure 5 shows that inside a radius of ${\sim}$ 0.1 pc, the electron density exceeds the critical density of 7 ${\times}$ 10$^5$ cm$^{-3}$
for collisional de-excitation of the ${^1D_2}$ level of O$^{2+}$. 
Third, the lower panel in Figure 5 shows a rapid increase in electron temperature inside the region where H is fully ionized. Cloudy predicts that the electron temperature exceeds 10${^7}$ K at the Balmer reverberation radius. Such a rapid rise in temperature correlates with an equally rapid decline
in the H${\alpha}$ emission line emissivity. This phenomenon leads to a central void, visualized in Figure 6, inside of which there is no Balmer line emission. The perimeter of this central void coincides with the Balmer reverberation radius. Thus, the Balmer reverberation radius 
appears to be just the inner radius of a larger volume of ionized gas that is producing Balmer line emission. The reverberating gas, identified with the inner ring of points in Figure 6, represents 15\% of the total, based
on the same percentage of the total Balmer emission line flux that is observed to be time variable, according to the {\it F${\rm_{var}}$} statistic \citep{Den09,Den10}. Although labelled variously, that statistic is routinely used to quantify variability amplitude by providing a measure of the fractional excess variance in the emission line flux
\cite[e.g.][]{Edl00}.

NGC 3516 is the third LLAGN
following NGC 3227 \citep{Dev13} and NGC 4051 \citep{DH13} for which the inner radius of the volume emitting the Balmer emission lines coincides with the Balmer reverberation lag. Various measures of BLR size in NGC 3516
are provided in the next section.

\begin{figure}
\includegraphics[width=84mm,clip]{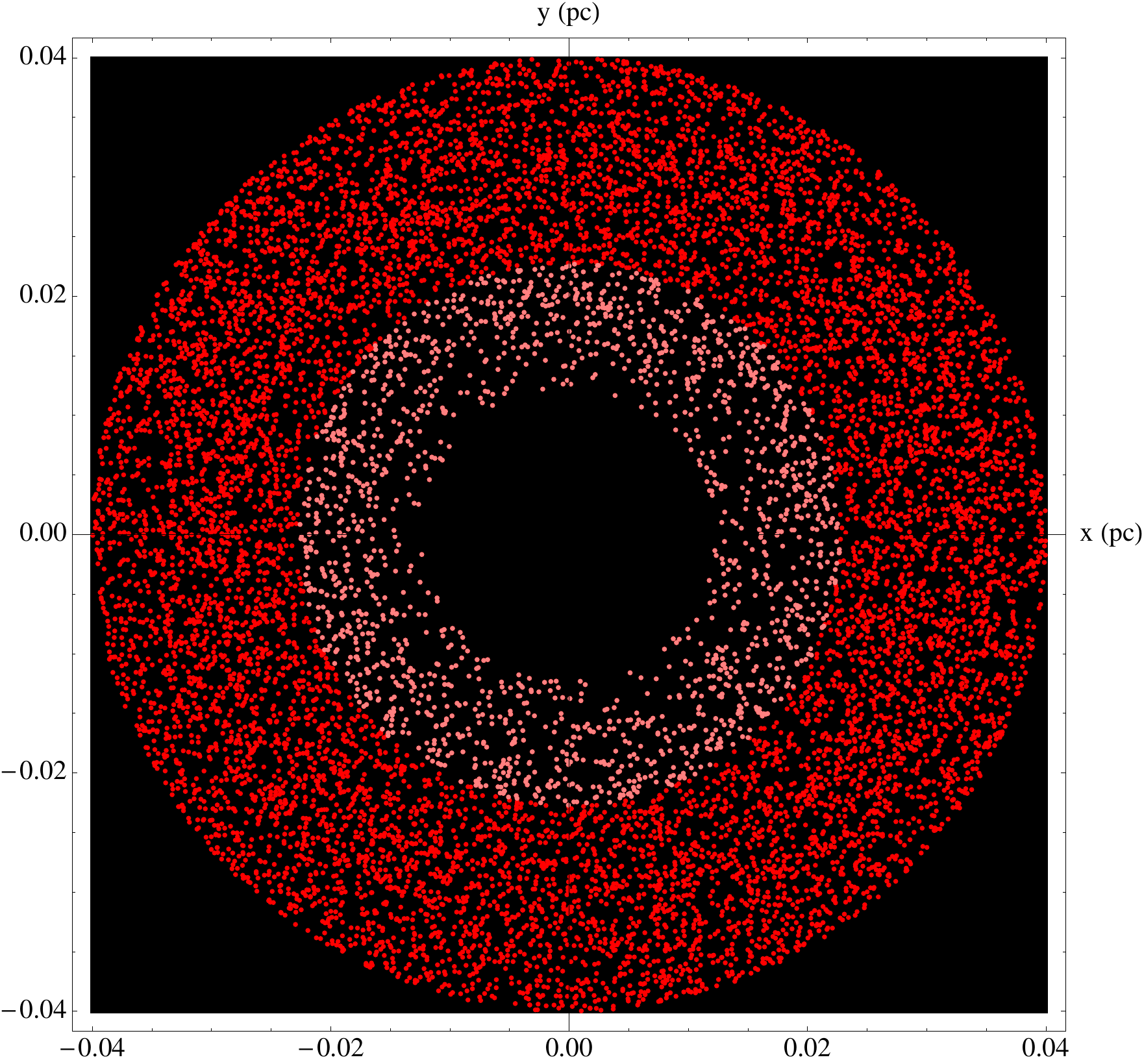}
\caption{Visualization of the BLR in NGC 3516. The units are pc for both the ordinate and abscissa.
The figure depicts the H${\alpha}$ emission line emissivity inside the dust reverberation radius measured by \cite{Kos14}. 
The number density of darker red dots is proportional to the H${\alpha}$ emission line emissivity. The inner lighter pink dots represent 15\% of the total that are reverberating \citep{Den10}. The central void identifies the region occupied by the ${\sim}$ 10${^7}$ K X-ray emitting plasma.}

\end{figure}

\subsection{BLR Size Estimates}

Knowing both the H${\alpha}$ emission line emissivity, and the central BH mass allows one to construct 
a model H${\alpha}$ emission line profile, an example of which is illustrated in Figure 7.
The line profile fitting method for estimating the size of the region producing broad Balmer line emission
has been described previously \citep{Dev11}. Briefly, the method employs a Monte Carlo simulation of 
a spherically symmetric distribution of ${\sim}$ 10$^4$ particles of light, the radial distribution of which is described by the H${\alpha}$ emission line emissivity predicted by Cloudy (see Figure 5).

Time resolved spectra discussed by \cite{Den09} indicate an infall component to the BLR in NGC 3516.
However, since the time variable component of the H${\beta}$ emission line represents only about ${\sim}$ 15\% of the total line flux, it is difficult to judge whether this observation is representative of the kinematic state of BLR as a whole. Nevertheless, for the purposes of computing the model line profile, every particle is assumed to be moving under the influence of gravity, and in free--fall according to the familiar equation v(r) = $\rm {\sqrt{ 2 G M_{\bullet}/r}}$, where v is velocity, G is the gravitational constant, M$_{\bullet}$ is the BH mass, and r is the distance of each point from the central supermassive BH. Such spherically symmetric free-fall models produce single peak broad Balmer emission line profile shapes. Discrete particle models also have the advantage that they reproduce the small-scale structure seen in broad emission line profiles, which is caused by random clumping in radial velocity space, as noted previously by \cite{Cap81}. 

The central mass determines the relationship between velocity and radius for each point of light and the emissivity determines the number of points at each radius. In the context of the inflow model there are two free parameters
available to model the line shape and they are the inner and outer radius of the emitting volume. The inner radius defines the full velocity width at zero intensity of the model broad emission line, and the outer radius defines the maximum intensity of the model broad emission line at zero velocity. Thus, 
comparing a normalized version of the observed broad emission line with the model one effectively constrains the inner and outer radii of the emitting volume using chi-squared minimization. 
For a BH mass of
31.7${^{+2.8}_{-4.2}}$ ${\times}$ 10$^{6}$ M${_{\sun}}$ \citep{Den10}
one finds that the inner radius, r${_i}$, of the region emitting the Balmer emission lines is 4 ${^{+1}_{-1}}$ l.d. which, within the uncertainties, is comparable to the Balmer reverberation lag, ${\tau_{peak}}$ = 7 ${^{+2}_{-1}}$ l.d., measured\footnotemark[3] by \cite{Den10}. \footnotetext[3]{For NGC 3516, there is a significant difference between ${\tau_{peak}}$, and ${\tau_{cent}}$, but there are more independent measurements that point to a time lag of ${\sim}$ 7 l.d. See \cite{Den10} for details.} 
Whereas the outer radius, r${_o}$, of the region emitting the Balmer emission lines is 47 ${^{+16}_{-16}}$ l.d. which coincides with the smallest of the dust reverberation radii measured by \cite{Kos14}.

The H${\alpha}$ emissivity predicted by Cloudy does a reasonably good job at reproducing the overall shape of the observed H${\alpha}$ emission line as illustrated in Figure 7, 
although there are some differences in detail. 
By design the model H${\alpha}$ emission line is symmetric about zero velocity, whereas the observed profile is obviously not. Additionally, the model does not explain the high velocity wings seen in the STIS spectra, suggesting a less precipitous decrease in the H${\alpha}$ emissivity at small radii than predicted by Cloudy. 
Nevertheless, the success of the Cloudy model is that the H${\alpha}$ emitting region, defined above, 
can explain both the H${\alpha}$ emission line profile shape (see Figure 7) and the Balmer emission line ratios (see Table 3) but it underestimates the extinction corrected H${\beta}$ luminosity by ${\sim}$ 60\%. Since the Balmer emission line emissivity is spatially extended (see Figure 5)
the model emission line luminosity can be increased to the observed value 
by increasing the outer radius 
to a  {\it luminosity} radius of 112 l.d., although that change causes the model Balmer emission line profile shape to deviate more from the observed one, and the model Balmer emission line ratios to no longer agree
with the extinction corrected values listed in Table 4. In summary,
there are several measures of BLR size and they include the Balmer reverberation radius, the dust reverberation
radius, the inner and outer radius of the volume required to explain the shape and relative intensities of the Balmer emission lines,
and lastly, the Balmer luminosity radius. These various size estimates are illustrated in Figure 5.

\section{Discussion}

Collectively the STIS observations constrain a Cloudy model for the BLR in NGC 3516 that consists of ${\sim}$ 500 M${_{\sun}}$ of dust-free H gas that is free-falling towards the central BH at a steady-state rate of ${\sim}$ 1 M${_{\sun}}$/yr.  
Even assuming radiatively inefficient accretion \citep[e.g.][]{Mer03}, the bolometric luminosity measured for this LLAGN \citep{Vas09} indicates 
that no more than 2\% of the inflowing material reaches the event horizon of the BH \citep{Bar09}
raising the question {\it where does the majority of the inflowing mass go?} Evidently, the inflow is diverted into an outflow. The mass outflow rate estimated for NGC 3516 by \cite{Bar09} accounts for only about 5\% of the inflowing mass quoted above. However, this discrepancy could be easily reconciled if the gas density in the outflow were about a factor of 20 higher than \cite{Bar09} assumed. Then mass would be conserved 
since the outflow rate would  
be similar to the inflow rate. A mechanism 
that would allow such an efficient redirection of matter most likely involves a magnetohydrodynamic
process since thermal energy and BH spin appear to be insufficient 
\cite[e.g.][]{Akt15}. 
A few other puzzles concerning NGC 3516 are discussed in the following.

\begin{figure}
 \includegraphics[width=84mm,clip]{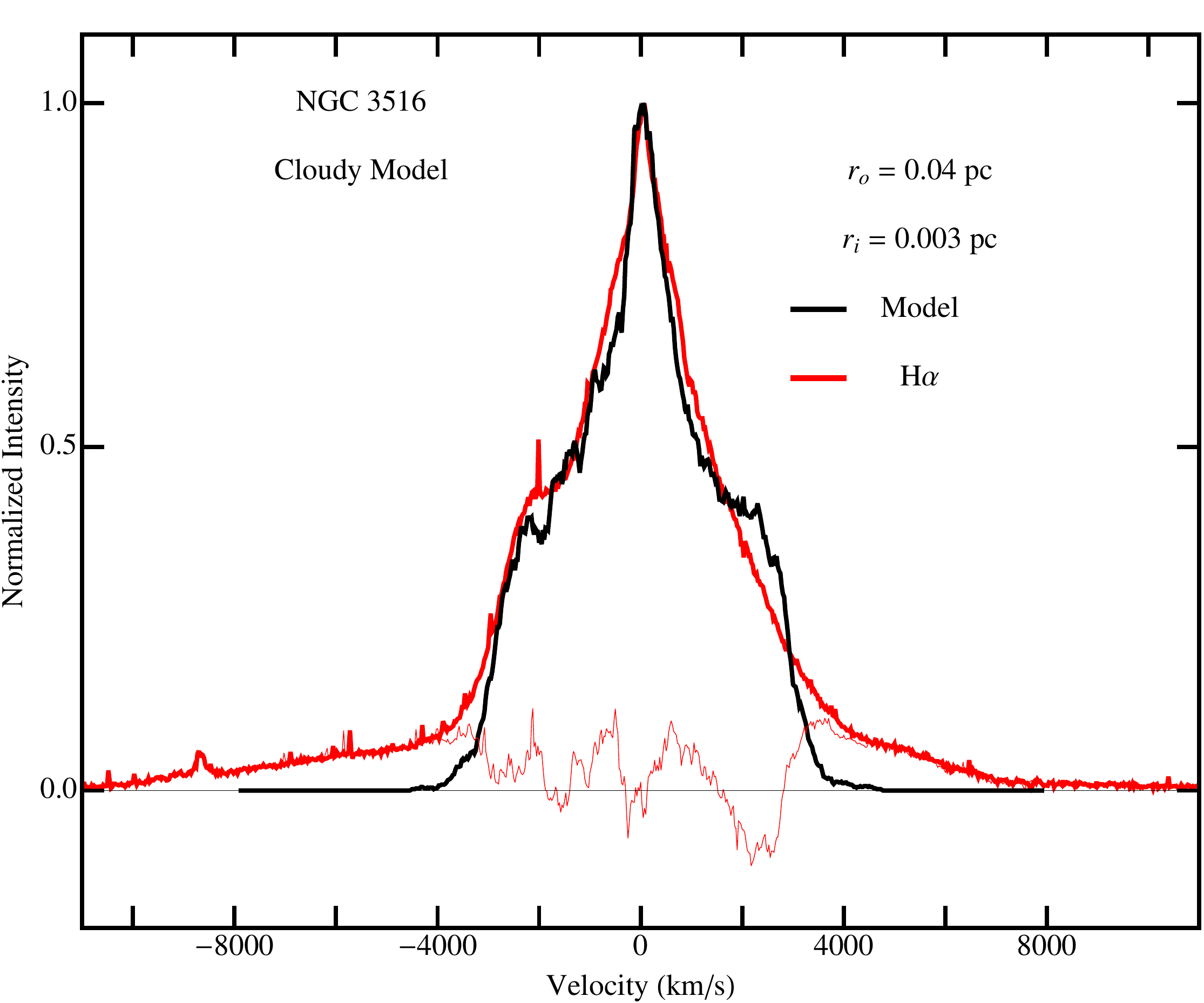}
 \caption{Cloudy model representation of the H${\alpha}$ emission line profile shape (black line) produced by the region depicted in Figure 6. The red line represents the observed normalized H${\alpha}$ emission line profile. The residual between the observed, and
model line, is represented by the thinner red line. }
\end{figure}

\subsection{X-ray Warm Absorber, Ionization Parameter, and H Column Density}

Cloudy predicts that inside the Balmer reverberation radius the H gas is an 10${^7}$ K plasma 
producing no H lines at all because the primary source of opacity is electron scattering. This inner sanctum is where the X-ray 
emission originates. Thus, according to this model, the X-ray, and Balmer emission are mutually exclusive, which would naturally explain the discordance between the time variability of these two types of radiation \citep{Edl00}.
Furthermore, according to this picture, the X-rays would have to pass through the 
ionized H to reach the observer which could explain the X-ray absorption features at ${\sim}$ 1 keV
described by \cite{Net02}. In fact, the properties of the {\it warm absorber} constrained by \cite{Net02}; a thin shell with an electron density ${\ge}$ 2.4 ${\times}$ 10${^6}$ cm${^{-3}}$, an electron temperature ${\sim}$ 3.5 ${\times}$ 10${^4}$ K, and a radius ${\le}$ 0.2 pc, 
almost perfectly describe the physical properties of the BLR gas illustrated in Figure 5. The obvious implication being that
the BLR ${\it is}$ the X-ray warm absorber. Subsequently, 
\citet[][and references therein]{Hue14}
have identified several {\it warm absorbers} covering a range of ionization parameter, {\it U(r)},
similar to the range predicted by Cloudy, as illustrated in Figure 5. However, \citet{Hue14} advocate {\it U(r)} increasing with radius
which is completely opposite to the dependence predicted by Cloudy (See Figure 5).
Furthermore, integrating the neutral H column of the Cloudy model over the entire range of radii depicted in Figure 5 leads to a H column density ${\sim}$ 2 ${\times}$10$^{24}$ atoms/cm$^{-2}$, which is 
an order of magnitude larger than estimated for any of the {\it warm absorbers} 
described by \citet[][and references therein]{Hue14}. Consequently, it is difficult to associate
any of the absorbers identified by \citet{Hue14} with the BLR gas.

\subsection{Forbidden Emission Lines}

What is visually striking about the spectra obtained with STIS of NGC  3516 is how faint the 
forbidden emission lines are compared to the H Balmer emission lines. For example, the 
observed [O\,{\sc iii}]/H${\beta}$, [O\,{\sc i}]/H${\alpha}$, [N\,{\sc ii}]/H${\alpha}$ and [S\,{\sc ii}]/H${\alpha}$ emission line ratios are so small that they
render NGC 3516 unclassifiable according to the diagnostic diagrams of \cite{Kew06}. However, when one compares the observed emission line ratios to the intrinsic ones predicted by the photoionization code Cloudy, none of the forbidden lines cited above are expected to be very bright except [O\,{\sc iii}].
Cloudy predicts the forbidden [O\,{\sc iii}] emission lines to be about one order of magnitude brighter than observed, even though the model electron density exceeds the critical density for collisional de-excitation of the ${^1D_2}$ level of O$^{2+}$ as mentioned previously in 
Section 3.2. Thus, if the Cloudy calculation is to be believed, then something is diminishing the brightness of the [O\,{\sc iii}] emission lines seen in NGC 3516. 

Looking at Figure 5, one possibility is dust obscuration. According to the results presented in the upper panel of Figure 5, dust could selectively obscure emission lines produced by any of the first four ionization stages of O. Furthermore, although not shown in the figure, the ionization gradient for O is similar to that of other ions including C, N, and S, because all these elements have similar ionization potentials. Thus, forbidden lines from those elements may also be obscured. Given that the dust extinction is virtually negligible to the H located inside the dust reverberation radius, any dust would have to be distributed in a face-on ring,
or annulus, in order to selectively affect just the forbidden lines. Such a geometry envisaged for the dust is reminiscent of a torus which is the basis for a unified model of AGN \cite[][and references therein]{Net15}.

Ideally, one would like to use Cloudy
to model the impact of dust on the forbidden line emission. Unfortunately, Cloudy has a serious limitation in that it has not reliably predicted {\it emergent emission line intensities} for all versions of the code including, and predating v13.02. Although not widely publicized, an admission to this effect is documented on the Cloudy simulations wiki hosted by Yahoo Groups\footnotemark[5] in a series of e-mail exchanges\footnotemark[6] at the end of the year 2014.  

\footnotetext[5]{${\rm https://groups.yahoo.com/neo/groups/cloudy\_simulations/info}$}
\footnotetext[6]{For example, message numbers; 2504, 2501, 2485, 2481}

According to the Cloudy documentation, the emergent line intensities include the radiative transfer effects involving dust beyond the region where the various emission lines are formed. Thus, the bug is related to the inclusion of dust in the Cloudy models. Regretably dust
is included in all models by default unless the user specifies {\it no grains} to disable it. A recent comparison of photoionization codes \citep{Peq01} did not address the inclusion of dust which is perhaps why this problem has gone unnoticed for so long.

The other half of the standard output generated by Cloudy titled {\it Intrinsic line intensities} is apparently unaffected by the bug, and it is those results that are used in this paper. However, according to the Cloudy documentation, the intrinsic line intensities do not include the radiative transfer effects involving dust beyond the region where the various emission lines are formed. Consequently the intrinsic line intensities are inappropriate for interpreting observed emission line spectra, unless the dust extinction to each region emitting each emission line is known a priori, and corrected for. Of special concern in this regard are several oft cited, and consequently influential papers dealing with spectroscopy of AGN that do not employ a dust extinction correction beyond the Galactic value. Collectively, several hundred papers utilizing Cloudy have been published in the professional literature. However, since it is not customary among the authors of those papers to declare which output they have been using, be it {\it intrinsic} or, {\it emergent}, or whether or not the {\it no grains} command was implemented, the reliability of any of the results presented is difficult to judge. This all underscores the pitfall associated with a discipline that relies almost entirely on a single photoionization code, in this case Cloudy. 
Having said all this, Cloudy intrinsic line intensities ${\it may}$ be useful for interpreting the visible Balmer emission line spectrum of NGC 3516, because for this particular AGN, it has been demonstrated in Sections 2.1 and 2.2, that the visible dust extinction is virtually negligible to the H in the BLR. 

\subsection{Low Metallicity}

If dust is not responsible for the weak forbidden [O\,{\sc iii}] emission lines observed for NGC 3516, then the only alternative is low metallicity. A metallicity that is a factor of 10 lower than the ISM value
causes Cloudy to reproduce the observed [O\,{\sc iii}]/H${\beta}$ ratio shown in Table 3. Metal poor gas suggests an origin in the circumgalactic medium. Perhaps the inflow, that we perceive as the BLR in NGC 3516, is just the terminus of a much larger inflow that originates from outside the galaxy. Such inflows of metal poor gas appear to be commonplace, observed in our own Galaxy and others \citep{L13}, but this is perhaps the first suggestion of an association between the BLR of an AGN, and a low metallicity accretion flow from the circumgalactic medium. Such inflows could also explain the low duty cycle observed for AGN activity in the local universe. 

\subsection{A UV--Visible Dichotomy?}

The main feature of the model presented here to explain the {\it visible} emission line spectrum of NGC 3516 is photoionization of {\it low} density gas ${\leq}$ 10$^8$ cm$^{-3}$ which leads to a spatially extended nebula surrounding the central UV--X-ray source. In contrast, \cite{Goa99b} use Cloudy to explain the UV spectrum of NGC 3516, in terms of photoionization of an ensemble\footnotemark[4] 
\footnotetext[4]{the LOC model \citep{B95}}
of optically thick {\it broad line clouds} with {\it high} density, 10$^9$ cm$^{-3}$ ${\leq}$ n${_e}$ ${\leq}$ 10$^{11}$ cm$^{-3}$.  These two models are mutually exclusive. The differences could be reconciled if 
there are two photoionization mechanisms at work, one in the visible, and one in the UV.
In effect, a UV--visible dichotomy whereby the UV emission lines are produced
by the accretion disk, and the visible emission lines from the photoionized nebula surrounding it.
As noted previously in Section 2.1 the largest variance in the visible Balmer emission line flux occurs in the line wings. However, the converse is true for the Ly${\alpha}$ emission line where the largest variance occurs in the line core \citep{Goa99b}. This distinction, if confirmed, would establish 
a basis for further investigation. 

Cloudy predicts the photoionized nebula will produce Ly${\alpha}$, 
and C\,{\sc iv} ${\lambda}$1542 emission lines, in addition to the visible lines
already mentioned (see Section 3.1). However, according to this model the nebular C\,{\sc iv} ${\lambda}$1542 emission would occur
in the vicinity of the dust reverberation radius, is quite likely attenuated by dust extinction, and 
is sufficiently distant from the central BH that it is expected to contribute only to the narrow component of  C\,{\sc iv} ${\lambda}$1542 discussed by \cite{Goa99b}. However, the nebula is expected to contribute significantly to the observed broad Ly${\alpha}$ emission, although it is difficult to explore further 
to what extent the nebula lines contribute in the UV given the shortcomings with Cloudy explained previously in Section 4.2.

\section{Conclusions}

A model has been presented which explains the relative intensities of the H${\alpha}$, H${\beta}$, and H${\gamma}$, emission lines in terms of a spatially extended, spherically symmetric distribution of neutral H gas that is photoionized by the central UV--X-ray source.  
Photoionization modeling with Cloudy indicates that the H${\alpha}$/H${\beta}$ emission line ratio is
a proxy for gas density, and constrains the neutral H density, ${\rho}$, to be log${_{10}}$${\rho}$(cm${^{-3}}$) = 7.4 at the Balmer reverberation radius. 
Collectively, the observations support a model, suggested previously by \cite{Net93},
in which the central UV--X-ray source is able to sublimate dust from a sizeable volume of H gas, permitting 
it to be photoionized. Modeling with the photoionization code Cloudy yields the following insights.
First, the Balmer emission line emissivity is essentially zero inside the Balmer reverberation radius. Thus, the Balmer reverberation radius marks the perimeter of a central cavity inside of which there is no Balmer
emission providing a natural explanation for the finite width observed for the Balmer emission lines. Second, the H gas is totally ionized between the Balmer reverberation radius and the dust reverberation radius.  That same H gas is associated with
an H${\alpha}$ emissivity that reproduces the overall shape of the observed H${\alpha}$ emission line
expected for gas in free-fall.
The Cloudy model further predicts forbidden [O\,{\sc iii}] emission lines that are one order of magnitude brighter than observed. The discrepancy may indicate that the
observed [O\,{\sc iii}] emission lines are attenuated by dust, or that the photoionized gas is of 
low metallicity, or both. A problem with the emergent line intensities computed by the Cloudy photoionization code precludes further investigation of this particular observation. 

\section*{Acknowledgments}

The author is grateful to Dr. Ranjan Vasudevan for providing the ionizing continuum illustrated in Figure 4, and to Dr. Hagai Netzer for helpful comments on an earlier draft of this manuscript. Special thanks also to Drs. Steven Willner, Jon Haass, Giorgio Lanzuisi and an anonymous referee for important
contributions that greatly influenced the outcome of this paper.

{\it Facilities:}  \facility{HST (STIS)}

\end{document}